\documentclass[a4paper]{jpconf}
\usepackage{anysize,graphicx,color}
\usepackage[ footnotesize, bf, margin=7pt, tableposition=top]{caption}
\usepackage{float}
\usepackage{sidecap}
\usepackage{wrapfig}
\usepackage[square, sort&compress]{natbib}
\begin{document}
\fontsize{10}{7}
\title{NaCo/SAM observations of sources at the Galactic Center}

\author{J. Sanchez-Bermudez$^1$, R. Sch\"odel$^1$, A. Alberdi$^1$, J. U. Pott$^2$}
\address{$^1$ Instituto de Astrof\'isica de Andaluc\'ia (CSIC), C/
  Glorieta de la Astronom\'ia S/N, 18008 Granada, Spain}
\address{$^2$ I. Physikalisches Institut, University of Cologne, Z\"ulpicher Str. 77, 50937 K\"oln, Germany}
\ead{joel@iaa.es, rainer@iaa.es, antxon@iaa.es, pott@ph1.uni-koeln.de}

\begin{abstract}
Sparse aperture masking (SAM) interferometry combined with Adaptive
Optics (AO) is a technique that is uniquely suited to investigate
structures near the diffraction limit of large telescopes. The strengths of the technique are a robust calibration of the Point Spread Function (PSF)
while maintaining a relatively high dynamic range. We used SAM+AO
observations to investigate the circumstellar environment of several
bright sources with infrared excess in the central parsec of the
Galaxy. For our observations, unstable atmospheric conditions as well
as significant residuals after the background subtraction presented
serious problems for the standard approach of calibrating SAM data via
interspersed observations of reference stars. We circumvented these
difficulties by constructing a synthesized calibrator directly from sources within the field-of-view. When observing crowded fields, this novel method can boost the efficiency of SAM observations because it renders interspersed calibrator observations unnecessary. Here, we presented the first NaCo/SAM images reconstructed using this method.
\end{abstract}

\vspace{-6mm}
\section{Introductory Remarks}

\vspace{3mm}
\begin{wrapfigure}{r}{0.52\textwidth}
\vspace{-16mm}
    \begin{center}
    \includegraphics[width=0.49\textwidth]{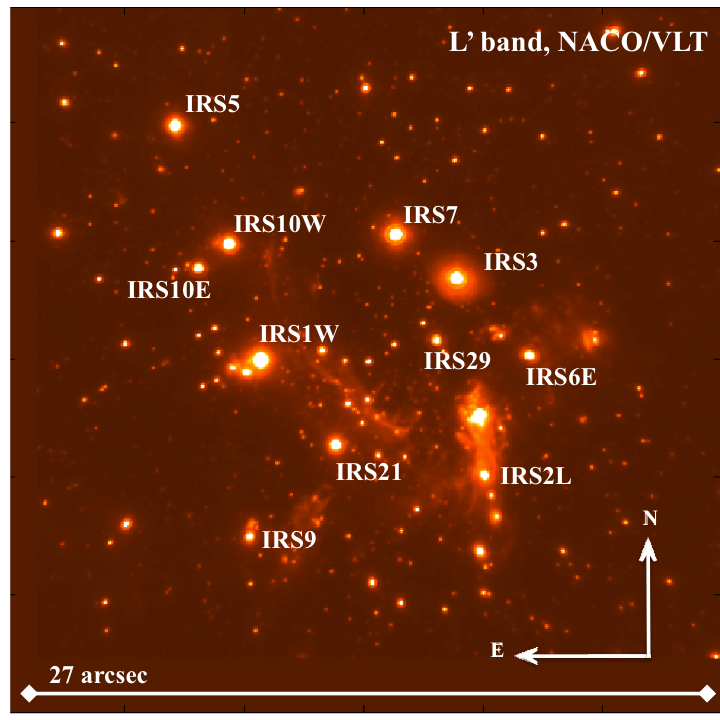}
\end{center}
\vspace{-8pt}
    \caption{The inner central parsec of the GC in the L'-
      band with NaCo/VLT. Our targets, the brighter massive stars in the field with known diffuse (resolved and/or unresolved) emission, are identified by their names.}
\vspace{-15pt}
\end{wrapfigure}

 \quad The Galactic center (GC) nuclear stellar cluster (NSC), the densest region of the
Milky Way ($\rho$ $>$ 10$^6$ M$_\odot$ pc$^{-3}$) \cite{Genzel2003ApJ, Schoedel2007AA}, contains the supermassive black hole (BH) SagittariusA*
(SgrA*) at its center \cite{Schoedel2003ApJ, Ghez2003ANS}  and exhibits signs of
at least two epochs of recent star formation \cite{Krabbe1995ApJ,
  Figer2004ASPC}.\ Theoretical considerations as well as observational
evidence suggest that the GC environment favors a top-heavy initial
mass function (IMF) \cite{Morris1993ApJ, Paumard2006ApJ,
  Bartko2009ApJ} (see however Do et al., these proceedings).\ On the order of 100 high-mass stars were created in the most recent ($\approx$4x10$^6$
years-old) starburst episode in the central parsec of the
GC. Since massive stars are rare in the Galaxy, this makes the Milky Way NSC an important site to study the properties of such objects (e.g.\ morphology, multiplicity) and their interactions with the interstellar medium (ISM). 

\vspace{9mm}
Figure 1 provides an overview of the central parsec of the Galaxy in
the form of an L'-band AO image obtained with the NaCo instrument of the
Very Large Telescope (VLT). Our targets, the brighter sources (4 $<$
mag$_L$ $<$ 7)  in the field of view (FOV),  are labeled with their
names. The principal selection criteria for this sample are that all of the sources (a) exhibit significant
mid-infrared excess, (b) are related to recent starburst events
($\sim$6 and $\sim$100 Myr ago) and (c) show extended structure like
dusty outflows and bow-shocks. Our sources can be divided into three
main groups: \textbf{(a) Evolved (super-) giants}, like IRS 7, 9, 3; \textbf{(b) Massive,
  post-main sequence stars,} such as IRS 6E, 29NE1 and \textbf{(c)
  Bow-shocks,} as IRS 1W, 21, 10W, and 5. 

\vspace{3mm}
\textit{The importance of a detailed study of such objects with high
  angular resolution
lies in the possibility to understand the role of
binarity/multiplicity in the dust formation processes, as well as the
circumstellar dust distribution}. In this context, Sparse Aperture Masking interferometry + Adaptive
Optics is one of the most suitable techniques to address these
goals.  Nevertheless, because of the difficulties we encountered during
the observations (bad atmospheric conditions, difficulty of measuring
the sky emission), the analysis of our SAM data requires the
 development of dedicated software and new reduction techniques to
 boost the efficiency of the observations. In this work, we
present a novel
calibration technique, as well as our first
reconstructed SAM images of the targets described above.

\section {Brief description of the technique}

\vspace{3mm}
\quad \textit{Adaptive Optics (AO)} is a powerful technique which is used to overcome
the degrading effects (mainly induced by the atmospheric turbulence) in the
wavefront (WF) of a specific source, while keeping high sensitivity.\ Nevertheless, in spite of the great promise
of this technique to provide us with diffraction limited images, it
still faces a
number of problems that make a precise and stable PSF calibration very
difficult.\ Such problems are, for example:\ (a) The PSF of the AO
systems is variable on different
timescales and highly sensitive to the observing
conditions (e.g. seeing, airmass, etc);\ (b) the AO performance is constrained by technological
limitations of the system (e.g.\ the number of sensors and actuators
that measure and correct the incoming WF); and (c) the
WF of the guide star is not the same as the one of the science
objects.\ Therefore, their wavefronts suffer different distortions because they take slightly different paths through the atmosphere (anisoplanatism) \cite{NACO_MANUAL,Tuthill2006SPIE}.

\vspace{3mm}
\textit{Sparse Aperture Masking (SAM) interferometry} is a technique which
 transforms a single dish telescope into a non-redundant Fizeau
 interferometer by  placing a mask with many holes in the pupil plane of the telescope
camera.\ The main advantage of SAM interferometry is the fact that the
non-redundant mask (NRM) removes most of the incoherent noise
produced by the atmospheric turbulence.\ In this way, a very well
defined and stable PSF can be obtained.\ The PSF resembles the
diffraction pattern of the mask (see Fig. 2). The nominal angular
resolution ($\theta$, which corresponds to the Full-Width Half-Maximum
of the beam) achieved by
this technique is of the order of $\theta$$\approx$$\lambda$/2D.\ This
is a factor of $\sim$2 better than the resolution obtained in standard imaging.\ Nevertheless, due to the fact that the NRM covers most of the
telescope pupil, SAM
suffers from low photon efficiency and is limited to bright objects (e.g.\
in the case of the VLT the magnitudes of the targets range between 4 to 12,
depending on the mask used). \textbf{A combination of the technologies of SAM and AO  offers us a unique tool
to recover high fidelity images and, hence, study structures between
one and few times the PSF size (e.g. in the case of the VLT, to analyze
morphologies in the regime of 50 to 150 mas), with the
strengths of the robust PSF calibration provided by SAM, and the
higher dynamic range obtained by the AO systems \cite{Tuthill2006SPIE}. } 

\vspace{0mm}
\begin{figure}[htp]
    \centering
    \includegraphics[width=10.7cm]{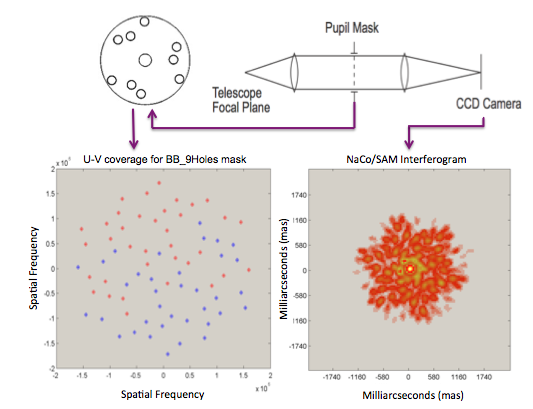}
    \caption{Sparse Aperture Masking layout. In the figure a NRM with nine holes is located in the
      pupil of the telescope. It is thus transformed into an
      interferometer with a non-redundant u-v coverage (lower-left gray scheme); note that the
    number of points sampled in the Fourier space correspond to all
    baselines formed by all the possible pair of holes combinations
    plus their complex conjugates (in this case, 36 red dots + 36 blue
    dots). The lower-right gray scheme shows the SAM 
   interferogram (i. e. interferogram) of a PSF
    obtained at the camera detector, with a shape that only depends on the geometry of the
  interferometric array. The upper part of the scheme was adapted from
the NaCo manual and Tuthill et al. (2006).}
    \label{SAM}
\end{figure}

\vspace{10mm}
\section {NaCo/SAM L-band observations of the Galactic Center }

\subsection{NaCo/SAM observational setup}

\begin{wrapfigure}{r}{0.52\textwidth}
\vspace{-15mm}
    \begin{center}
    \includegraphics[width=0.49\textwidth]{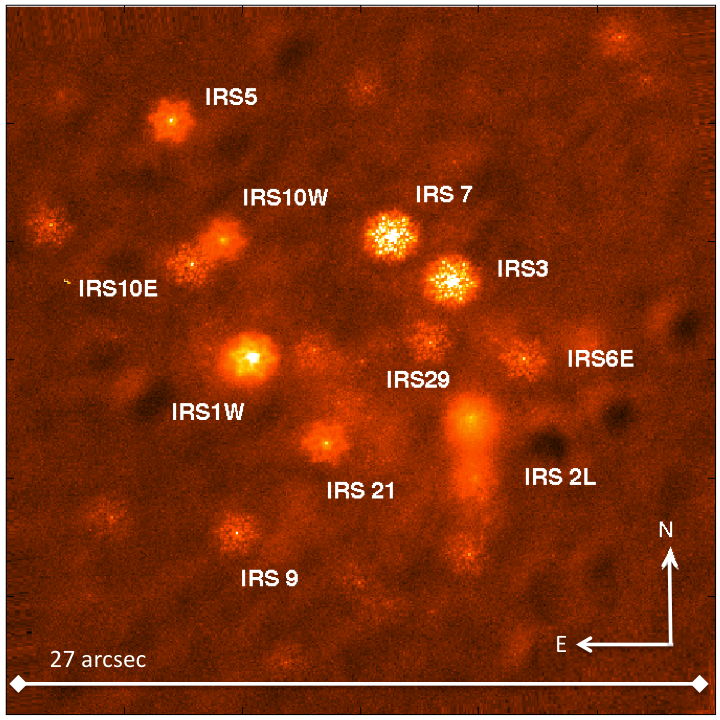}
\end{center}
\vspace{ -2mm}
    \caption{NaCo/SAM
  FOV. The main targets of our sample show clearly the interference
  pattern of the NRM overimposed on them and are labeled with their
  names. Since SAM is a Fizeau type interferometry, it keeps the
  entire FOV of the camera used.}
\vspace{-5mm}
\end{wrapfigure}

\vspace{3 mm}
In June 2010 we performed VLT NaCo/SAM observations of the central
parsec of the GC NSC in the L band, using two different filters; (i) L'
(3.8 $\mu$m ) and  (ii) NB 3.74 (3.74 $\mu$m). The
observations in the L' filter were conducted using the
\textit{BB\_9holes} mask, while the ones in the NB 3.74 filter were taken using the
\textit{9\_holes} mask. The main reason to use these NRMs is because
they are optimized to be utilized with broad band
and narrow band filters, respectively. The
observations were recorded using the Pupil-Tracking mode of  NaCo (which freezes the pupil and makes the FOV  rotate in
opposite direction to the parallactic angle). The so-called {\it cube
  mode} was used, which saves each single data integration time (DIT)
frame. The advantage of the cube mode is that it allows the selection
of the best frames. Frame selection can become necessary when AO
performance is not stable, as it was the case for the observations described here. Our observations consisted of a group
of 13 sets of data-cubes composed of 20 - 30 exposures with 5 sec of
DIT. Because SAM removes the low
spatial frequencies related to the atmospheric noise, chopping
techniques were not used to eliminate the near-infrared background
radiation. Nevertheless, since the GC NSC is extremely crowded, separate
sky observations were taken on a dark cloud a few arcmin to the
north-west  of Sgra*. IRS7, the brightest target in our FOV,
was used as guide star for the  the IR WFS of NaCo. Figure 3 shows a single Naco/SAM exposure with our
targets identified on it.

\subsection{Data Reduction}

\vspace{1mm}
\quad The reduction process of our NaCo/SAM data-cubes is divided into two principal steps:
(i) standard imaging reduction and (ii) 
interferometric analysis. 

\subsubsection{Standard Imaging:}
In this reduction part, as in any other optical-NIR imaging
observations, all data-cubes were flat-fielded, bad-pixel corrected and
background subtracted. Depending on the observational data set, the sky
template for the background subtraction was obtained by calculating
either (i) the median of the sky
cubes or (ii) through the median of the lower values at each pixel from the dithered images, when no dedicated sky observations close in time were available.  As our images were taken using the Pupil
Tracking mode of NaCo, all cubes had to
be derotated to correctly compute the interferometric
observables. All data-cubes were frame selected,
discarding bad images through the analysis of the cube flux
statistics ($\sim$ 70-90 \% of the frames  were eliminated). Finally,
each one of the individual targets was stored into a subcube of 128 x 128 pixels. 

\subsubsection{Interferometric reduction process:}
To obtain the interferometric observables (squared
visibilities and closure phases) we used Michael Ireland's IDL
pipeline. This code calculates the Fourier transform of all the
frames in a given data-cube and
convolve them with a \textit{matched filter} template, which is the Fourier
transform of all baselines formed by the holes in the
NRM. Due to the facts that (i) each hole in the NRM has a finite
radius, and (ii) that our data cubes are stored into a discrete array of pixels, the normalized flux at every baseline in the Fourier space is spread into a
group of pixels, or \textbf{splodge}, around the central spatial
frequency of the used baseline (see Fig 4). Therefore, every raw (not calibrated) squared
visibility is calculated by the squared value of the sum of the
normalized flux at a given splodge,
divided by the total flux received (i.e.\ baseline zero). The closure phases are the arguments of the bispectra, i.e.\ the ensemble of all visibility products for non-redundant baseline triangles. It is important to remark
that: the visibility amplitudes give us information about
the compactness and extension of the source brightness distribution in all the spatial frequencies
sampled by the baselines of our interferometer, while closure phases
contains information on the orientation and symmetry of the
target structure and are robust to any observational effect (e.g.\ atmospheric turbulence).

\vspace{3 mm}
In order to increase the u-v coverage and to get a better sampling of the sources brightness
distribution (see Fig. 5)  we use the Earth rotation synthesis through the combination of the visibilities
obtained by observations of the targets at different parallactic angles. An example of the IRS7  raw 
visibilities  and closure phases is shown in Figure 6.

\subsubsection{Calibration of visibilities and closure phases.} 
The standard observing technique for NaCo/SAM is the observation of a
science target interwoven with observations of a nearby point-source
reference object (i.e.\ the calibrator). This star should be
observed at similar airmass and in the same AO configuration as the
science target.\ Therefore, calibrator sources should be of similar
brightness as the targets.\ Unfortunately, this was not possible for
our observations because of the following difficulties:

\vspace{3mm}
(a) Rapidly changing observing
conditions (clouds, highly variable seeing, etc.) affected the
calibrator observations, so that it was hard to obtain data of similar
quality on source and calibrators; 

(b) The GC is highly extincted.\ Hence, it is very difficult to find calibrators that are
sufficiently bright in the L band ($\sim$ 4 mag$_L$) because the ones that fulfill this
condition are too bright in the K-band so that they would saturate the
infrared wavefront sensor of NaCo. Therefore, the observed calibrators  were
systematically fainter than our brighter targets (e.g. IRS7 and IRS
3);

(c) The observations were affected by strong residual patterns that remained in the images after background subtraction. The reason for
these residuals are not clear, but they are typical for
L-band observations, where the background fluctuates rapidly (see
Fig. 7).

\vspace{-3mm}
\begin{figure}[htp]
\begin{minipage}[ht]{9cm}
\begin{center}
\hspace{-5mm}
\includegraphics[width=9cm,clip]{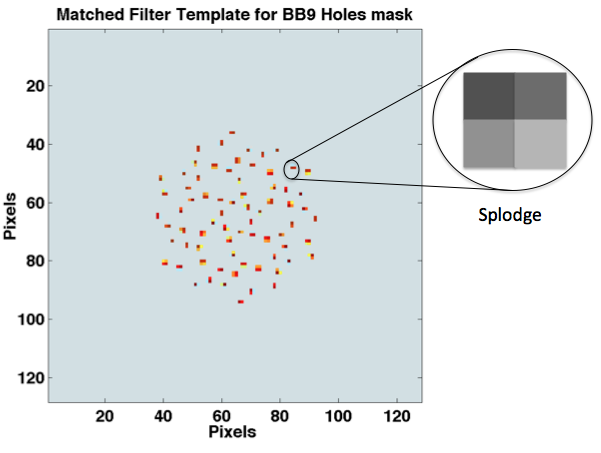}
\caption[Short caption for figure 1]{\label{labelFig4} Matched Filter
  Template. The total flux at every spatial frequency (given by 
  every baseline of the mask)  sampled in the matched filter is spread
  in a series of pixels called ``splodges''. The total amount of
  pixels which form a splodge is determined by the size of the mask holes.}
\end{center}
\end{minipage}
\hspace{-4mm}
\begin{minipage}[ht]{7.5cm}
\begin{center}
\includegraphics[width=7.5cm,clip]{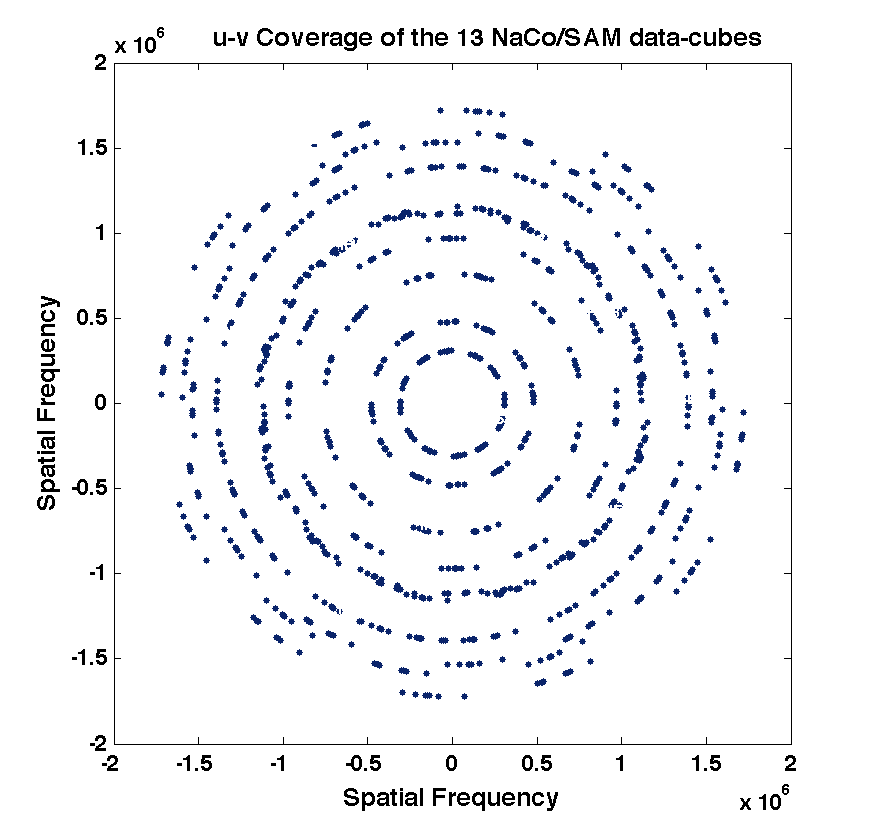}
\vspace{-7mm}
\caption[Short caption for figure 2]{\label{labelFig5} NaCo/SAM total u-v
  coverage. The total u-v coverage of our interferometer is produced
  using the Earth Rotation Synthesis through the combination of all
  data-cubes visibilities. The beam is symmetric in all directions.}
\end{center}
\end{minipage}
\end{figure}

\begin{figure}
\vspace{-12mm}
\begin{minipage}[ht]{8.7cm}
\begin{center}
\includegraphics[width=8.7cm,clip]{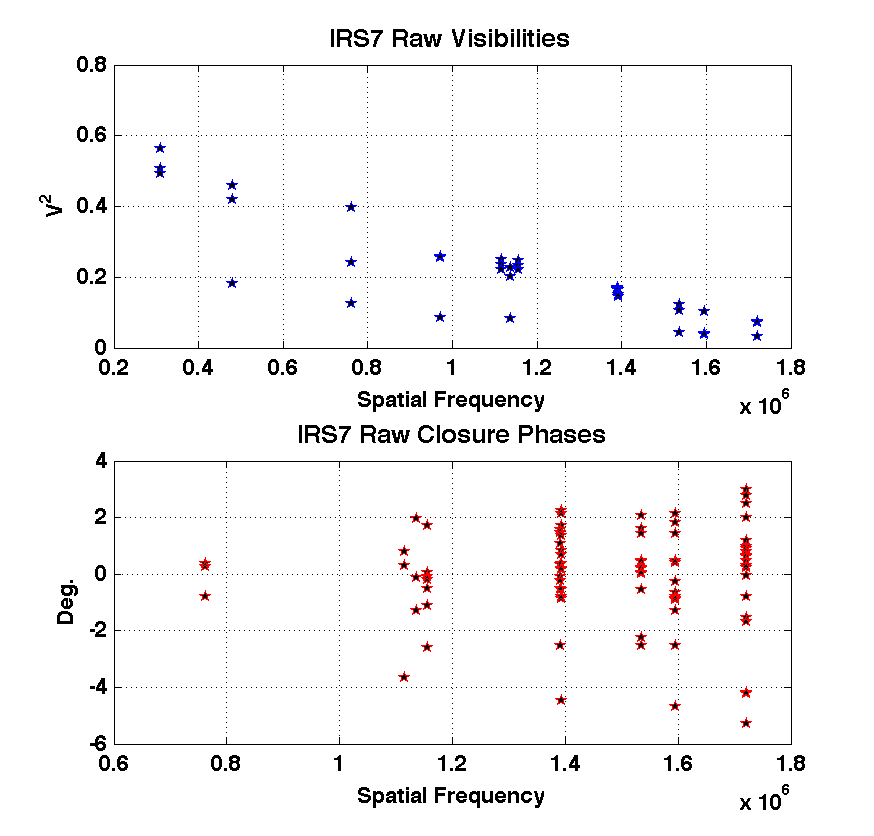}
\caption[Short caption for figure 1]{\label{labelFig6} Raw Squared
  visibilities and Closure phases of IRS7. On the figure we have the
  uncalibrated 36 visibilities and 84 closure phases of one of the
  observed IRS7 L-band data-cubes using the \textit{BB\_9holes} mask.}
\end{center}
\end{minipage}
\begin{minipage}[ht]{7cm}
\begin{center}
\includegraphics[width=7cm,clip]{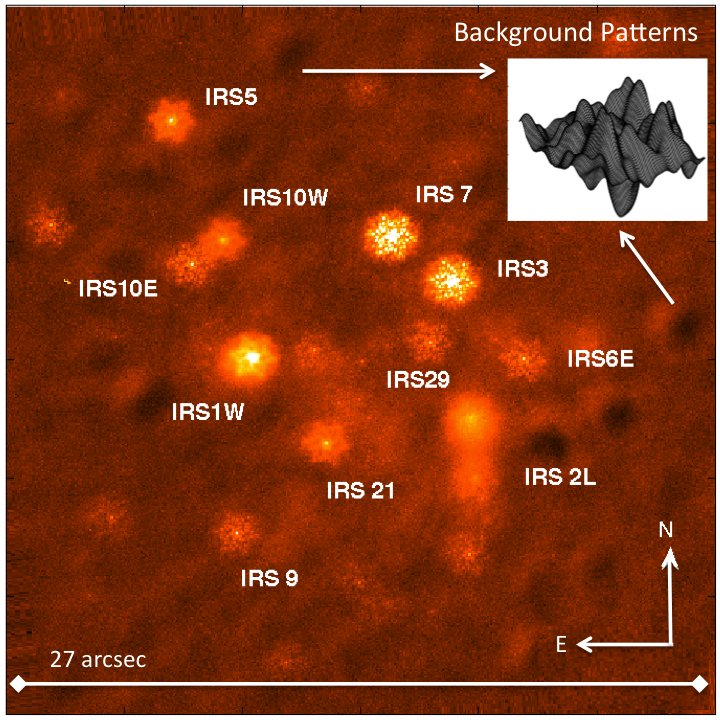}
\caption[Short caption for figure 1]{\label{labelFig7} L-band NaCo/SAM
  background patterns. All frames of our observations exhibit some
  background patterns spread all over the FOV. The angular scale
  of the background variations is of the order of the size of the PSF
  ($\sim$ 1").}
\end{center}
\end{minipage}
\end{figure} 

\vspace{110 mm}
In order to overcome these problems we developed a calibration
technique, which we called the {\it ``Synthetic Calibrator''}. This
method takes advantage of the presence of several bright sources in our FOV and creates a PSF by source superposition, using StarFinder \cite{Diolaiti2000} routines. The images of stars are extracted from the FOV, cleaned from the contamination of secondary sources, locally
background subtracted, centered with  sub-pixel accuracy, normalized
and combined by a median superposition. The resulting PSF is thus obtained directly from the field, without the need to observe a calibrator source. Some of the advantages are: (a) No interwoven
observations of standard calibrators are needed,
therefore, the observing conditions are exactly the same for both the
calibrator and the target; (b)
the calibrators can be as bright as science targets.

\vspace{3 mm}
Note that, a priori, we do not know the intrinsic
source-structure of our targets. This can can be a source of
systematic uncertainties when extracting a synthetic
calibrator. Nevertheless, the latter effect, as well as the background
residuals, are effectively
minimized by the median super-position of several sources when
creating the synthetic calibrator, as our tests have confirmed. Figure
8 shows and example of the calibrated visibilities and closure phases
of two of our targets IRS 7 and IRS 1W obtained with this technique. Of course, the Synthetic Calibrator technique can only be used when several sufficiently bright sources are present within the field-of-view. It may thus be ideal for observations of stars in dense clusters.

\vspace{-2 mm}
\begin{figure}[htp]
\centering
\hspace{-3mm}
\includegraphics[width=15cm,clip]{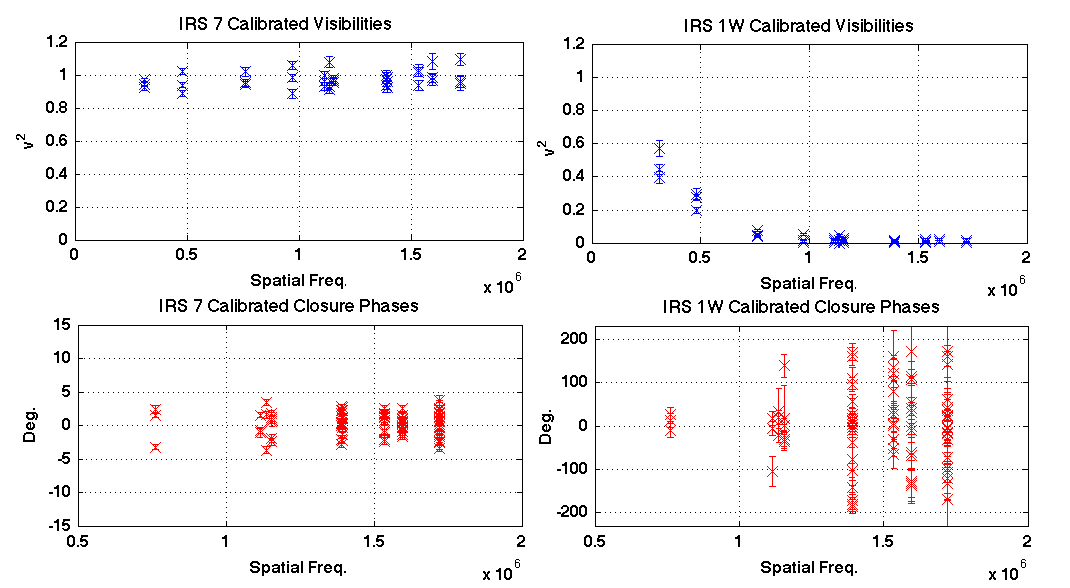}
\caption[Short caption for figure 1]{\label{labelFig8}  Calibrated
  Visibilities and Closure Phases of IRS 7 and IRS
  1W. Calibrated
  visibilities of IRS 7 are constant and close to the unit, demonstrating
  that  this source is a point-like object; its closure phases also
  confirm the point-symmetry of the object (left). On the other hand,
  the visibilities of IRS 1W suggest the existence of an extended
  structure larger than the interferometric beam; its closure phases
  demonstrate the point asymmetry of this source (right). }
\end{figure}

\vspace{-5 mm}
\section{Image reconstruction.}

\vspace{3 mm}
\quad Once the calibrated interferometric observables had been obtained, image reconstruction was done with the BSMEM
package \cite{Buscher1994}, which uses a maximum entropy algorithm to
reconstruct the interferometric maps. Figure 9 displays a comparison between
three of our reconstructed SAM images (IRS 1W, IRS 5 and IRS 10W) in the L-band (upper part) and the AO deconvolved
Ks images of
the same sources (lower part) obtained with the 10-meter Keck telescope by Tanner
et al.\ (2005) \cite{Tanner2005}. The resolution achieved with our SAM observations in
the L-band  (at 3.8 $\mu$m, $\theta$$\approx$60 mas) is similar to the one
obtained by the AO system of Keck telescope in the Ks band (at 2.2
$\mu$m, $\theta$$\approx$45 mas). This demonstrates that the nominal resolution of the
NaCo/SAM technique has been achieved. On the other hand, Figure 10
presents for comparison a set of
six L-band images taken with the VLT. The first column shows our SAM
images, the second one displays Lucy-Richardson \cite{Lucy1974AJ}
devoncolved images and the third one corresponds to raw AO
images. The resolution  and 
quality of the reconstructed SAM
images surpass the raw and deconvolved AO images at the same frequency.

\vspace{1 mm}
\begin{figure}[htp]
\begin{minipage}[ht]{15cm}
\begin{center}
\vspace{-4 mm}
\includegraphics[width=15cm,clip]{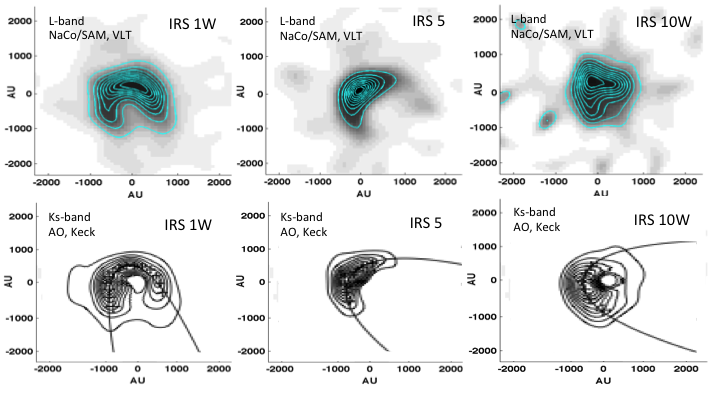}
\caption[Short caption for figure 1]{\label{labelFig9} NaCo/SAM vs AO
  Keck images. At the top, NaCo/SAM images of IRS 1W, IRS 5 and IRS 10W in the L-band; in the bottom, the same sources
 are
  presented as deconvolved AO images in te K-band. The angular
  resolution of both subsets of images is similar, even though there
  is a difference in wavelength. The K-band images are adapted from Tanner
  et al (2005). Contours of NaCo/SAM images represent 10, 20, 30 40, 50, 60,
70, 80 and 90 \% of the total normalized flux.}
\end{center}
\end{minipage}
\begin{minipage}[ht]{15cm}
\begin{center}
\vspace{5 mm}
\includegraphics[width=15cm,clip]{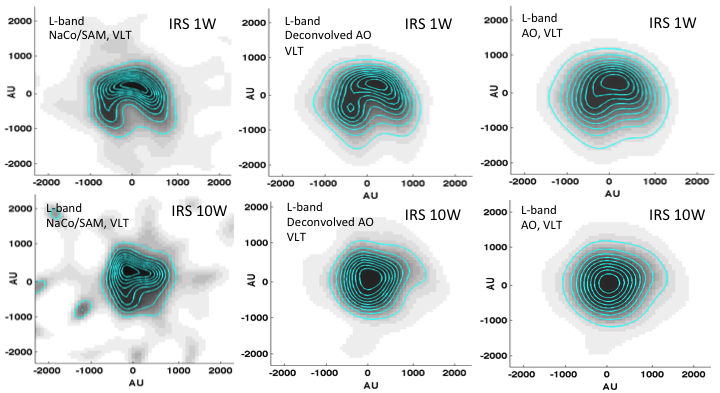}
\caption[Short caption for figure 2]{\label{labelFig10} NaCo/SAM vs
  AO. IRS 1W and IRS 10W are presented in three subset of images: at
  the left, in the form of NaCo/SAM images; in the middle, as
  Lucy-Richardson deconvolved images; to the right, as
  non-deconvolved AO images. Note how the resolution on the NaCo/SAM
  images overpass the ones obtained by the other two subsets.
  Contours represent 10, 20, 30 40, 50, 60,
70, 80 and 90 \% of the total normalized flux. }
\end{center}
\end{minipage}
\end{figure}

\vspace{3 mm}
\section{Summary}

\vspace{3 mm}
\quad We overcome the difficulties of our NaCo/SAM observations
(highly variable weather conditions, background instabilities and
calibration problems) with
the creation of  \textit{``The Synthetic Calibrator''} from the
median superposition of several sources in the field, which according
to our results appears to be a
promising new way to calibrate SAM observations of dense stellar fields. The strength of this technique
relies on the opportunity to observe multiple targets at the same time in Stellar
Clusters, thus increasing the efficiency of the observations
by eliminating the necessity of interspersed observations between standard
calibrators and targets.\ From our reconstructed NaCo/SAM images we can conclude that
the nominal L-band NaCo/SAM resolution was achieved ($\theta$$\approx$60 mas) and that
the quality of reconstructed maps quality clearly
exceeds Lucy-Richardson deconvolved images. With these results, the scientific analysis
of the sources, according to the scientific case described in the
introduction, can be addressed.

\ack{\small
JSB acknowledges support by the ``JAE-Pre'' programme of the Spanish Consejo
Superior de Investigaciones Cient\'ificas (CSIC). RS acknowledges support by the Ram\'on y Cajal programme, by grants
  AYA2010-17631 and and AYA2009-13036 of the Spanish Ministry of
  Science and Innovation, and by grant P08-TIC-4075 of the Junta de
  Andaluc\'ia. AA acknowledges support by grant AYA2009-13036 of the
  Spanish Ministry of Science and Innovation and by grant P08-TIC-4075
  of the Junta de Andaluc\'ia.}

{\small \bibliography{iopart-num}}

\end{document}